\documentclass[twocolumn,showpacs,preprintnumbers,amsmath,amssymb]{revtex4}
\usepackage{graphicx}
\usepackage{dcolumn}
\usepackage{bm}

\begin{document}

\title {Quantum Fourier Transform in a Decoherence-Free Subspace}

\author{Jian-wu Wu}
\email{wujw03@mails.tsinghua.edu.cn}
\author{Chun-wen Li}
 \email{lcw@tsinghua.edu.cn}
\author{Re-bing Wu}
\email{wurebing98@mails.tsinghua.edu.cn}

\affiliation{Department of Automation, Tsinghua University,
Beijing, 100084, P. R. China
\\
}

\date{\today}

\begin{abstract}
Quantum Fourier transform is of primary importance in many quantum
algorithms. In order to eliminate the destructive effects of
decoherence induced by couplings between the quantum system and
its environment, we propose a robust scheme for quantum Fourier
transform over the intrinsic decoherence-free subspaces. The
scheme is then applied to the circuit design of quantum Fourier
transform over quantum networks under collective decoherence. The
encoding efficiency and possible improvements are also discussed.
\end{abstract}


\maketitle

\section{\label{sec:level1}Introduction}

Quantum Fourier transform (QFT) plays essential roles in various
quantum algorithms such as Shor's algorithms \cite{QFT1,QFT2,QFT7}
and hidden subgroup problems \cite{QFT2,QFT8,QFT9}. Inspired by
the exponential speed-up of Shor's polynomial algorithm for
factorization \cite{QFT1}, many people investigated the problem of
efficient realization of QFT in a quantum computer
\cite{QFT3,QFT4,QFT5,QFT6,QFT7}. Up to now, many improvements have
been made. In \cite{QFT3}, Moore and Nilsson showed that QFT can
be parallelized to linear depth in a quantum network, and upper
bound of the circuit depth was obtained by Cleve and Watrous
\cite{QFT4} for computing QFT with a fixed error. In \cite{QFT5}
the actual time-cost for performing QFT in the quantum network was
examined. Further, Blais \cite{QFT6} designed an optimized quantum
network with respect to time-cost for QFT.

In practice, the decoherence problem induced by the unavoidable
coupling of quantum system with the environment have to be
considered in circuit design for QFT over a quantum network. If no
measure is taken, decoherence will destroy the encoded quantum
information. Many methods have been proposed to suppress
decoherence in a quantum system, among which, an important scheme
is to encode the quantum information into the decoherence-free
subspaces or subsystems (DFS)
\cite{DFS1,DFS2,DFS3,DFS4,DFS5,DFS6,DFS7,DFS8,DFS9} of quantum
system. Theoretically, DFSs are completely isolated from the
noises \cite{DFS1,DFS2,DFS3,DFS4}. A large amount of discussions
about DFS have appeared in the literature
\cite{DFS5,DFS6,DFS7,DFS8,DFS9}.

In this paper, we will take advantage of the decoherence-free
subspaces to develop a novel scheme for performing QFT in a
quantum computer. The circuits designed in this way have the
robustness against noise in the procedure of implementing QFT. The
paper is organized as follows: in Sec.II, some notations and the
preliminary knowledge on DFS and QFT will be reviewed; in Sec.III,
general method will be introduced for implementing QFT in DFS; in
sec.IV, circuits will be designed to perform QFT in the DFS of a
quantum network with respect to weak collective decoherence (WCD)
and strong collective decoherence (SCD) respectively; in Sec.V,
the efficiency of the circuit and possible improvements will be
discussed; finally, a conclusion will be made in Sec.VI.

\section{Notations and preliminaries}

Suppose the quantum system {\bf S} under consideration is coupled
to an environment {\bf E}. The overall system is governed by the
Hamiltonian in the form of \cite{DFS7}:

\noindent \begin{equation}\label{eq1} H_{SE} = \sum_{\alpha \in
\Lambda} S_\alpha \otimes E_{\alpha} ,\end{equation}

\noindent where $S_{\alpha}(E_{\alpha}),(\alpha \in \Lambda)$ are
operators acting on the state space
$\mathcal{H}_{S}(\mathcal{H}_{E})$ of {\bf S}({\bf E}), and the
index set $\Lambda$ contains all the possible couplings between
the system and the environment. Assume $S_{\alpha}(\alpha \in
\Lambda)$ span a $\dag$-closed associate algebra $\mathcal{A}$.
According to \cite{DFS10}, $\mathcal{A}$ is isomorphic to a direct
sum of $d_J \times d_J$ complex matrix algebras, each with
multiplicity $n_J$ \cite{DFS6,DFS7}

\noindent \begin{equation}\label{eq2} \mathcal{A} \cong
\bigoplus_{J \in \mathcal{J}} I_{n_J} \otimes
{\mathcal{M}(d_J,\mathcal{C}))} ,\end{equation}

\noindent where the index set $\mathcal{J}$ labels all the
irreducible components of $\mathcal{A}$. Correspondingly, the
system Hilbert space $\mathcal{H}_{S}$ can be decomposed into a
similar form

\noindent \begin{equation}\label{eq3} \mathcal{H}_{S} =
\bigoplus_{J \in \mathcal{J}} \mathcal{C}^{n_J} \otimes
{\mathcal{C}^{d_J}}.
\end{equation}

All the subsystem spaces $\mathcal{C}^{n_J}(J \in \mathcal{J})$ in
the right hand side of Eq.(\ref{eq3}) correspond to
decoherence-free subsystems of the quantum system {\bf S}.
Particularly, $\mathcal{C}^{n_J}$ gives a decoherence-free
subspace of the quantum system {\bf S} when $d_J=1$.

Quantum network under collective decoherence (CD) provides a nice
paradigm for the DFSs. Roughly speaking, all qubits of a quantum
network under CD are coupled to the environment in the same
manner. In the literature\cite{DFS7}, two types of CD, weak
collective decoherence (WCD) and strong collective decoherence
(SCD), are frequently discussed.

SCD is defined as the decoherence due to the interaction
Hamiltonian

\noindent \begin{equation}\label{eq4}
H_{SE}=\sum_{\alpha =x,y,z}
S_\alpha \otimes E_{\alpha},
\end{equation}

\noindent where $S_{\alpha}={\sum}_{i =1}^n \sigma_{\alpha}^i$,
and $\sigma_{\alpha}^i (\alpha=x,y,z)$ represents the Pauli matrix
$\sigma_{\alpha}$ that corresponds to the local operation on the
$i^{th}$ qubit.

If only one term appears in the right hand side of Eq.(4), i.e.
the system is coupled to the environment only in one direction,
the induced decoherence is called WCD. Without loss of generality,
the Hamiltonian can be written as

\noindent \begin{equation}\label{eq5} H_{SE} =  S_z \otimes E_z.
\end{equation}

Next, we give a brief description of QFT implemented over an
$n$-qubit quantum network. Mathematically, the quantum Fourier
transformation $QF_n$ can be expressed as \cite{QFT6}:

\noindent \begin{equation}\label{eq6} QF_n:
|\varphi\rangle\rightarrow \frac{1}{2^{\frac{n}{2}}}
\sum_{\phi=0}^{2^{n}-1} e^{i2\pi\varphi\phi} |\phi\rangle.
\end{equation}

Denote the state of the quantum network by the qubit string
$|s_ns_{n-1}\cdots s_2s_1\rangle, (s_t\in \{0,1\},t=1,2,\cdots,n)$
in which the $t^{th}$ qubit is at the state $|s_t\rangle$. The
transformation $QF_n$ can be realized by applying the following
sequence of quantum gates (all the gate sequences in this paper
are operated from the right to left one by one)

\noindent \begin{equation}\label{eq7} T^{(1)}T^{(2)}\cdots
T^{(n-1)}T^{(n)},
\end{equation}

\noindent where

\noindent \begin{equation}\label{eq8}
T^{(k)}=P^{(1,k)}(\frac{\pi}{2^{k-1}})P^{(2,k)}(\frac{\pi}{2^{k-2}})\cdots
P^{(k-1,k)}(\frac{\pi}{2})H^{(k)} .
\end{equation}

Eq.(\ref{eq8}) includes two classes of elementary quantum gates,
$H^{(k)}$ and $P^{(i,j)}(\theta)$. The local Hadamard gate
$H^{(k)}$ represents

\noindent \begin{equation}\label{eq9}
H=\frac{1}{\sqrt{2}}\left[\begin{array}{cc}1 & 1\\1 &
-1\end{array}\right]
\end{equation}

\noindent over $k^{th}$ qubit. The controlled-phase-shift gate
$P^{(i,j)}(\theta)$ represents the action

\noindent \begin{equation}\label{eq10}
P(\theta)=\left[\begin{array}{cccc}1&0&0&0\\
0&1&0&0\\0&0&1&0\\0&0&0&e^{i\theta}\end{array}\right]
\end{equation}

\noindent over $i^{th}$ (control) and $j^{th}$ (target) qubits.

In addition, there are three important elementary gates that will
be used in this paper. The controlled-NOT gate $CN^{(i,j)}$ flips
the $j^{th}$ qubit (target qubit) when the $i^{th}$ qubit (control
qubit) is at the state $|1\rangle$, and nothing is done when the
control qubit is at the state $|0\rangle$. The rotation gate
$R^{(k)}(\alpha)$ realizes the unitary transformation over
$k^{th}$ qubit:

\noindent\begin{eqnarray}\label{eq11}& & {|0\rangle\longrightarrow
\cos\alpha |0\rangle+\sin\alpha
|1\rangle},\nonumber\\
& &{|1\rangle\longrightarrow -\sin\alpha |0\rangle+\cos\alpha
|1\rangle}.
\end{eqnarray}

\noindent The controlled-rotation gate $CR^{(i,j)}(\beta)$
realizes the rotation $R^{(j)}(\beta)$ over $j^{th}$ qubit (target
qubit) when the state on the $i^{th}$ qubit (control qubit) is
$|1\rangle$, and nothing is done when the state on the $i^{th}$
qubit (control qubit) is $|0\rangle$.

More details about DFS and QFT can be found in ref.
\cite{QFT6,QFT7,DFS6,DFS7} and the references therein.

\section{Scheme for performing QFT in a DFS}

In the following parts of this paper, we focus the study on
implementing QFT over decoherence-free subspaces. If not claimed,
the abbreviation DFS will indicate only the decoherence-free
subspace. Suppose $\mathcal{H}_{DFS}$ is an $n_J$ dimensional DFS
of the quantum system {\bf S}, then one can select
$2^{[\log_2n_J]}$ orthonormal states in $\mathcal{H}_{DFS}$ to
construct at most $[\log_2n_J]$ new qubits. For clarity, we call
these qubits logical-qubits, and the original qubits
physical-qubits.

Next, we will discuss how to realize a robust QFT algorithm over
these logical-qubits. Similar to (\ref{eq6}), we define the QFT
over a DFS by

\noindent \begin{equation}\label{eq12} {QF_n}_{(DFS)}:
|\hat{\varphi}\rangle\rightarrow \frac{1}{2^{\frac{n}{2}}}
\sum_{\phi=0}^{2^{n}-1} e^{i2\pi\varphi\phi}
|\hat{\phi}\rangle,\end{equation}

\noindent where the states
$|\hat{\phi}\rangle=|\hat{s}_n\hat{s}_{n-1}\cdots\hat{s}_1\rangle,\hat{s}_t\in\{\hat{0},\hat{1}\},t=1,2,\cdots,n$
are basis states of the logical-qubits.

The basic idea for realizing ${QF_n}_{(DFS)}$ is as follows.
Notice that the two classes of gates $H^{(k)}$ and
$P^{(i,j)}(\theta)$ play the central roles in the QFT
\cite{QFT4,QFT5,QFT6,QFT7}, we will construct correspondingly two
similar classes of quantum gates for implementing QFT in a DFS,
denoted by one-logical-qubit gate $H_{DFS}^{(k)}$ acting on the
$k^{th}$ logical-qubit and two-logical-qubit gate
$P_{DFS}^{(i,j)}(\theta)$ acting on the $i^{th}$ and $j^{th}$
logical-qubits respectively. These two classes of gates should
fulfil two requirements: one is that they are invariant operators
on the state space of the logical-qubits; the other is that they
operate on the logical-qubits in the same manner as $H^{(k)}$ and
$P^{(i,j)}(\theta)$ on the physical-qubits.

Similar to the general QFT described in Eq.(\ref{eq6}), we can
realize ${QF_n}_{(DFS)}$ by applying the following sequence of
quantum gates

\noindent \begin{equation}\label{eq13}
T_{DFS}^{(1)}T_{DFS}^{(2)}\cdots T_{DFS}^{(n-1)}T_{DFS}^{(n)},
\end{equation}

\noindent where

\noindent \begin{equation}\label{eq14}
T_{DFS}^{(k)}=P_{DFS}^{(1,k)}(\frac{\pi}{2^{k-1}})P_{DFS}^{(2,k)}(\frac{\pi}{2^{k-2}})\cdots
P_{DFS}^{(k-1,k)}(\frac{\pi}{2})H_{DFS}^{(k)} .
\end{equation}

\noindent Gate sequence (\ref{eq13}) provides us the general
strategy for designing a circuit to implement QFT over the DFS in
a quantum system.

Concretely, let $|\hat{l}\rangle,l=0,1,\cdots,n_J-1$ be $n_J$
orthonormal states in the DFS $\mathcal{H}_{DFS}$, and
$|\hat{l}\rangle,l=n_J,n_J+1,\cdots,2^n-1$ be $2^n-n_J$
orthonormal states in the orthogonal complementary space of
$\mathcal{H}_{DFS}$. Then between $|\hat{l}\rangle$ and the
natural basis $|s_ns_{n-1}\cdots
s_2s_1\rangle,s_t\in\{0,1\},t=1,2,\cdots,n$ of space
$\mathcal{H}_S$ there exists an unitary transformation $U$, i.e.

\noindent \begin{equation}\label{eq15} |s_ns_{n-1}\cdots
s_2s_1\rangle=U|\hat{l}\rangle,
\end{equation}

\noindent where

\noindent \begin{equation}\label{eq16}
l=s_n\cdot2^{n-1}+s_{n-1}\cdot2^{n-2}+\cdots+s_1\cdot2^0.
\end{equation}

Let $m$ be an integer no greater than $\log_2{n_J}$. Here we
choose $|\hat{l}\rangle,l=0,1,\cdots,2^m-1$ to construct $m$
logical-qubits for performing m-qubit QFT over the DFS, and
rewrite them as

\noindent \begin{equation}\label{eq17}
|\hat{l}\rangle=|\hat{s}_m\hat{s}_{m-1}\cdots\hat{s}_1\rangle,\hat{s}_t\in\{\hat{0},\hat{1}\},t=1,2,\cdots,m,
\end{equation}

\noindent where

\noindent \begin{equation}\label{eq18}
l=s_m\cdot2^{m-1}+s_{m-1}\cdot2^{m-2}+\cdots+s_1\cdot2^0.
\end{equation}

\noindent Then

\noindent \begin{eqnarray}\label{eq19}
& &U^{-1}H^{(k)}U|\hat{s}_m\cdots\hat{s}_{k+1}\hat{0}\hat{s}_{k-1}\cdots\hat{s}_1\rangle\nonumber\\
&= &U^{-1}H^{(k)}|\underbrace{{0\cdots0}}_{n-m}s_m\cdots
s_{k+1}0s_{k-1}\cdots s_1\rangle\nonumber\\
&=&U^{-1}\frac{1}{\sqrt{2}}(|\underbrace{{0\cdots0}}_{n-m}s_m\cdots
s_{k+1}0s_{k-1}\cdots s_1\rangle\nonumber\\
& &+|\underbrace{{0\cdots0}}_{n-m}s_m\cdots s_{k+1}1s_{k-1}\cdots
s_1\rangle)\nonumber\\
&=&\frac{1}{\sqrt{2}}(|\hat{s}_m\cdots\hat{s}_{k+1}\hat{0}\hat{s}_{k-1}\cdots\hat{s}_1\rangle\nonumber\\
&
&+|\hat{s}_m\cdots\hat{s}_{k+1}\hat{1}\hat{s}_{k-1}\cdots\hat{s}_1\rangle).
\end{eqnarray}

\noindent Similarly, we have

\noindent \begin{eqnarray}\label{eq20}
& &U^{-1}H^{(k)}U|\hat{s}_m\cdots\hat{s}_{k+1}\hat{1}\hat{s}_{k-1}\cdots\hat{s}_1\rangle\nonumber\\
&
=&\frac{1}{\sqrt{2}}(|\hat{s}_m\cdots\hat{s}_{k+1}\hat{0}\hat{s}_{k-1}\cdots\hat{s}_1\rangle\nonumber\\
&
&-|\hat{s}_m\cdots\hat{s}_{k+1}\hat{1}\hat{s}_{k-1}\cdots\hat{s}_1\rangle),
\end{eqnarray}

\noindent and

\noindent \begin{eqnarray}\label{eq21}
&&U^{-1}P^{(i,j)}(\theta)U|\hat{s}_m\hat{s}_{m-1}\cdots\hat{s}_1\rangle\nonumber\\
&=&\left\{\begin{array}{cc}
|\hat{s}_m\hat{s}_{m-1}\cdots\hat{s}_1\rangle,
&{\rm if}|\hat{s}_i\hat{s}_j\rangle\in\{|\hat{0}\hat{0}\rangle,|\hat{0}\hat{1}\rangle,|\hat{1}\hat{0}\rangle\}\\
e^{i\theta}|\hat{s}_m\hat{s}_{m-1}\cdots\hat{s}_1\rangle, &{\rm
if}\,\, |\hat{s}_i\hat{s}_j\rangle=|\hat{1}\hat{1}\rangle
\end{array}\right.,\nonumber\\
\end{eqnarray}

\noindent where $1\leq i\neq j\leq m$.

From Eqs.(\ref{eq19}-\ref{eq21}), we can see that if $U$ can be
constructed by elementary gates, then $U^{-1}H^{(k)}U$ and
$U^{-1}P^{(i,j)}U$ are feasible realizations for the two gates
$H_{DFS}^{(k)}$ and $P_{DFS}^{(i,j)}(\theta)$. Thus the
realization of the unitary transformation $U$ is crucial for
building the circuits to implement QFT in a DFS.

The remainder tasks, then, are to find the transformation $U$ in
Eq.(\ref{eq15}) and build a circuit to realize it. From the theory
of universal quantum computation\cite{QC1}, any unitary operator
can be constructed by a sequence of universal elementary gates. In
most cases it is not easy to obtain such explicit decompositions.
Whereas, as will be shown in the next section, it is possible to
build up a circuit for $QF_{n(DFS)}$ over the quantum network
under collective decoherence with a finite number of elementary
gates.

\section{Circuits for QFT over quantum networks under collective decoherence}
\subsection{\label{sec:level2}The weak collective decoherence case}

In the quantum networks under WCD, nontrivial DFS exists only when
the original network has no less than two physical-qubits
\cite{DFS7}. For the simplest case, the DFS in a two-qubit quantum
network under WCD is spanned by the orthonormal states
$|01\rangle$ and $|10\rangle$ \cite{DFS7}, with which one can
build up one logical-qubit, i.e.

\noindent \begin{equation}\label{eq22} |\hat{0}\rangle=|01\rangle,
\end{equation}

\noindent and

\noindent \begin{equation}\label{eq23} |\hat{1}\rangle=|10\rangle.
\end{equation}

For a $2n$-qubit quantum network under WCD, we use the orthonormal
states
$|\hat{s}_n\rangle\otimes|\hat{s}_{n-1}\rangle\otimes\cdots\otimes|\hat{s}_1\rangle,(\hat{s}_t\in\{\hat{0},\hat{1}\},t=1,2,\cdots,n)$,
where $|\hat{s}_{t}\rangle$ represents the $t^{th}$ logical-qubit
extracted from the $(2t-1)^{th}$ and $(2t)^{th}$ physical-qubits,
to construct the circuit for robust QFT. It can be verified that
all these states are contained in the biggest DFS.

\begin{figure}[h]
\centering
\includegraphics[totalheight=1.1in]{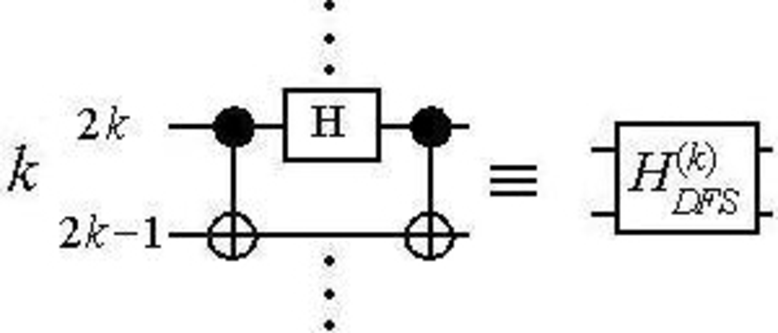}
\caption{Circuit for the gate $H_{DFS}^{(k)}$ in quantum network
under WCD. The element with $\bigoplus$ corresponds to a
controlled-NOT gate with control on the filled circle and target
on the $\bigoplus$. (In this paper, all the different
logical-qubits are labelled by numbers in the first column of the
figures, while the individual physical-qubits are labelled by the
numbers in the second column.)} \label{fig1}
\end{figure}

\begin{figure}[h]
\centering
\includegraphics[totalheight=1.85in]{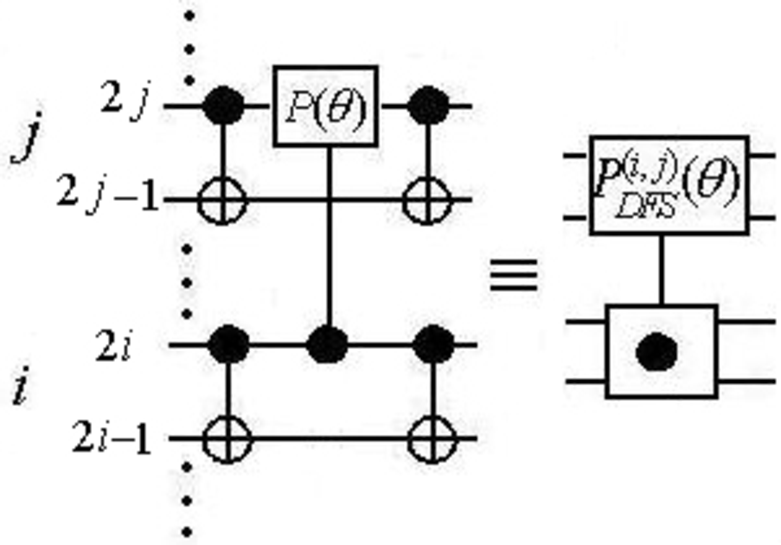}
\caption{Circuit for the controlled-phase-shift gate
$P_{DFS}^{(i,j)}(\theta)$ over the $i^{th}$ and $j^{tj}$
logical-qubits in quantum network under WCD.} \label{fig2}
\end{figure}

Over these logical-qubits, it is observed that $H_{DFS}^{(k)}$ and
$P_{DFS}^{(i,j)}(\theta)$ can be directly constructed from a
sequence of elementary gates as follows (the circuits are given in
Fig.\ref{fig1} and Fig.\ref{fig2}):

\noindent \begin{equation}\label{eq24}
H_{DFS}^{(k)}=CN^{(2k,2k-1)}H^{(2k)}CN^{(2k,2k-1)},
\end{equation}

\noindent and

\noindent \begin{eqnarray}\label{eq25}
P_{DFS}^{(i,j)}(\theta)&=&(CN^{(2i,2i-1)}CN^{(2j,2j-1)})P^{(2i,2j)}(\theta)\nonumber\\
&&\times(CN^{(2i,2i-1)}CN^{(2j,2j-1)}).
\end{eqnarray}

Being able to perform the gates $H_{DFS}^{(k)}$ and
$P_{DFS}^{(i,j)}(\theta)$ introduced above with elementary gates,
one can now integrate the circuit for an n-qubit QFT over a
2n-physical-qubit quantum network under WCD. The transformation
$QF_{n(DFS)}$ can be realized by replacing the $H_{DFS}^{(k)}$ and
$P_{DFS}^{(i,j)}(\theta)$ in the gate sequence (\ref{eq13}) with
those in Eqs.(\ref{eq24}) and (\ref{eq25}).

Let $U_n=CN^{(2,1)}CN^{(4,3)}\cdots CN^{(2n,2n-1)}$. Observing
that the term $CN^{(2t,2t-1)}$ commutes with $H^{(2k)}$ when
$t\neq k$ and commutes with $P^{(2i,2j)}(\theta)$ when $t\neq i$
or $j$, we have

\noindent \begin{equation}\label{eq26}
U_n^{-1}H^{(2k)}U_n=CN^{(2k,2k-1)}H^{(2k)}CN^{(2k,2k-1)}=H_{DFS}^{(k)},
\end{equation}

\noindent and

\noindent \begin{eqnarray}\label{eq27}
U_n^{-1}P^{(2i,2j)}(\theta)U_n&=&(CN^{(2i,2i-1)}CN^{(2j,2j-1)})P^{(2i,2j)}(\theta)\nonumber\\
& &\times(CN^{(2i,2i-1)}CN^{(2j,2j-1)})\nonumber\\
&=&P_{DFS}^{(i,j)}(\theta).
\end{eqnarray}

Therefore, we can choose $U_n$ as the unitary transformation $U$
in Eq.(\ref{eq15}):

\noindent \begin{equation}\label{eq28}
U=U_n=CN^{(2,1)}CN^{(4,3)}\cdots CN^{(2n,2n-1)}.
\end{equation}

\begin{figure}[h]
\centering
\includegraphics[totalheight=1.4in]{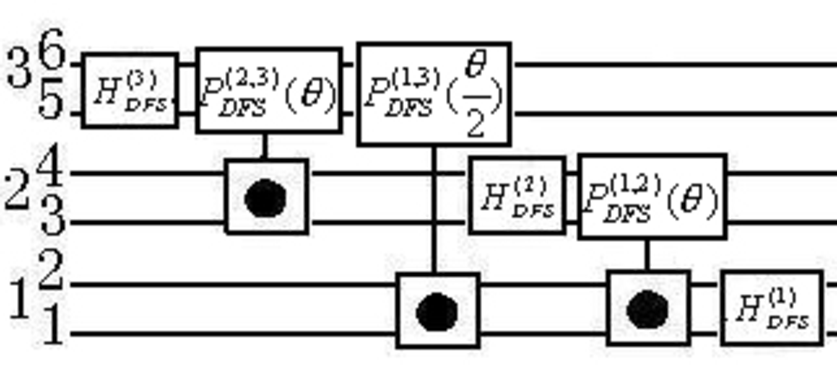}
\caption{Circuit for realizing three-qubit QFT on a
six-physical-qubit quantum network under WCD.
 $\theta=\frac{\pi}{2}$; the gates  $H_{DFS}^{(k)}$
and $P_{DFS}^{(i,j)}(\theta)$ are those given in Fig.1 and Fig.2.}
\label{fig3}
\end{figure}

Consider the three-qubit QFT as a simple example, the
transformation $QF_{3(DFS)}$ can be realized by applying
$H_{DFS}^{(k)}$ and $P_{DFS}^{(i,j)}(\theta)$  in the sequence as
follows (see the circuit in Fig.\ref{fig3}):

\noindent \begin{equation}\label{eq29}
H_{DFS}^{(1)}P_{DFS}^{(1,2)}(\frac{\pi}{2})H_{DFS}^{(2)}
P_{DFS}^{(1,3)}(\frac{\pi}{4})P_{DFS}^{(2,3)}(\frac{\pi}{2})H_{DFS}^{(3)}.
\end{equation}

\subsection{The strong collective decoherence case}

It is more complicated to design the circuit for QFT over quantum
networks under SCD than WCD. The corresponding condition for the
existence of a DFS is more critical. Quantum network with four
physical-qubits is of the smallest scale to ensure the existence
of a nontrivial DFS, which is spanned by two orthonormal states
\cite{DFS7}

\noindent \begin{equation}\label{eq30}
|\tilde{0}\rangle=\frac{1}{2}(|01\rangle-|10\rangle)(|01\rangle-|10\rangle),
\end{equation}

\noindent and

\noindent\begin{eqnarray}\label{eq31}
|\tilde{1}\rangle&=&\frac{1}{\sqrt{12}}(|01\rangle-|10\rangle)(|01\rangle-|10\rangle)\nonumber\\
&+&\frac{1}{\sqrt{3}}|0\rangle(|01\rangle-|10\rangle)|1\rangle \nonumber\\
&-&\frac{1}{\sqrt{3}}|1\rangle(|01\rangle-|10\rangle)|0\rangle.
\end{eqnarray}

Naturally, $|\tilde{0}\rangle$ and $|\tilde{1}\rangle$ form one
logical-qubit. By dividing the physical qubits into 4-qubit units,
one can use the canonical basis
$|\tilde{s}_{n}\rangle\otimes|\tilde{s}_{n-1}\rangle\otimes\cdots\otimes|\tilde{s}_{1}\rangle,(\tilde{s}_{t}\in\{\tilde{0},\tilde{1}\},t=1,2,\cdots,n)$
, where $|\tilde{s}_{t}\rangle$ represents the $t^{th}$
logical-qubit extracted from the $(4t-3)^{th}$ to $(4t)^{th}$
physical-qubits, to construct $n$ logical-qubits in a
$4n$-physical-qubit quantum network under SCD.

To perform QFT over the logical-qubits obtained above, it is still
crucial to design the circuits for the corresponding two classes
of gates $H_{DFS}^{(k)}$ and $P_{DFS}^{(i,j)}(\theta)$. Here we
directly give the form of unitary transformation $U$ in
Eq.(\ref{eq15}), then the gates $H_{DFS}^{(k)}$ and
$P_{DFS}^{(i,j)}(\theta)$ are obtained according to Sec.{III}.

\begin{figure}[h]\centering\includegraphics[totalheight=3.0in]{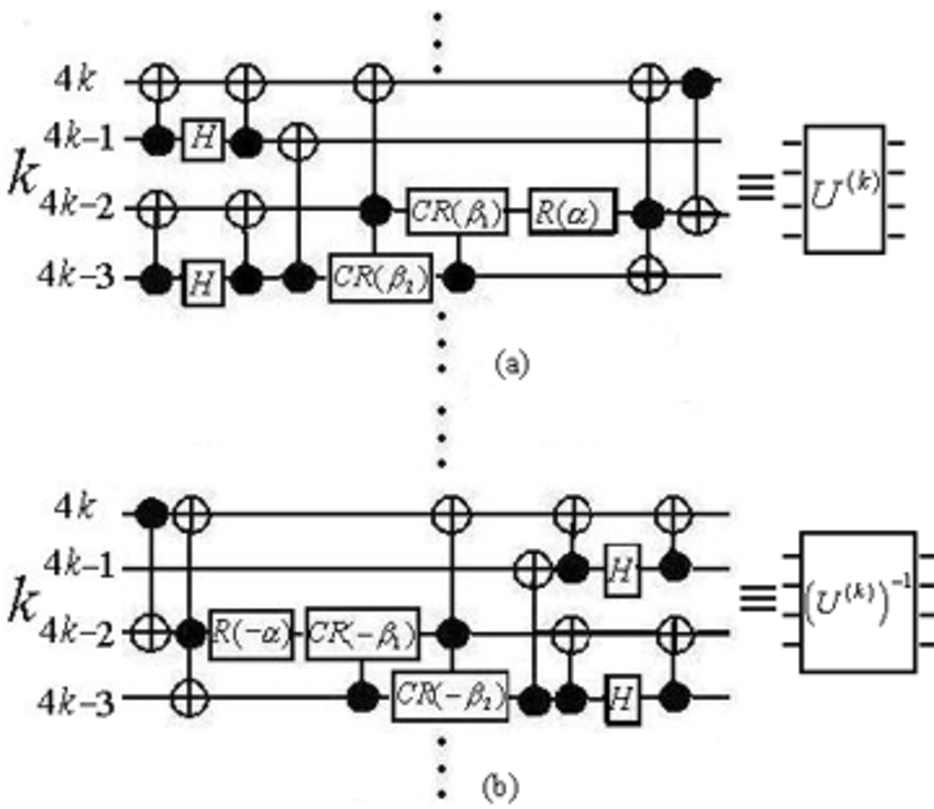}
\caption{(a) Circuit for the unitary transformation $U^{(k)}$
corresponding to the case that implementing QFT in a quantum
network under SCD. (b) Circuit for the inverse transformation
${U^{(k)}}^{-1}$.}\label{fig4}
\end{figure}

Let $U^{(k)}$ be an unitary transformation on the physical-qubits
from $(4k-3)^{th}$ to $(4k)^{th}$, which is realized by applying
the sequence of elementary gates as follows (see the circuits for
the transformation $U^{(k)}$ and its inverse in Fig.\ref{fig4}):

\noindent \begin{eqnarray}\label{eq32}
U^{(k)}&=&CN^{(4k,4k-2)}CN^{(4k-2,4k-3)}CN^{(4k-2,4k)}\nonumber\\
& &\times R^{(4k-2)}(\alpha)CR^{(4k-3,4k-2)}({\beta}_1)CR^{(4k-2,4k-3)}({\beta}_2)\nonumber\\
&&\times CN^{(4k-2,4k)}CN^{(4k-3,4k-1)}CN^{(4k-3,4k-2)}\nonumber\\
&&\times CN^{(4k-1,4k)}H^{(4k-3)}H^{(4k-1)}\nonumber\\
&&\times CN^{(4k-3,4k-2)}CN^{(4k-1,4k)},
\end{eqnarray}

\noindent where $\alpha=\pi-\arcsin\frac{1}{\sqrt{3}}$,
$\beta_1=-\pi+\arcsin\frac{1}{\sqrt{3}}$,
$\beta_2=-\frac{\pi}{4}$.

\noindent Then, in a $4n$-physical-qubit quantum network, one of
the feasible realization of the transformation $U$ is:

\noindent \begin{equation}\label{eq33} U=U^{(n)}U^{(n-1)}\cdots
U^{(1)}
\end{equation}

\begin{figure}[h]\centering\includegraphics[totalheight=1.4in]{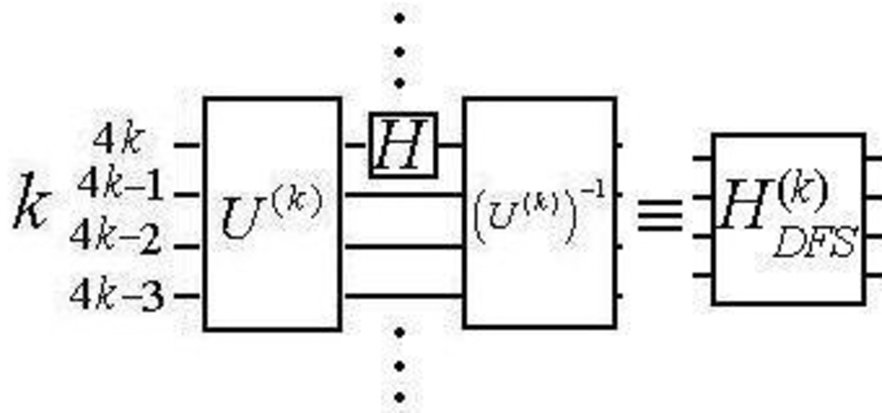}
\caption{Circuit for the Hadamard gate $H_{DFS}^{(k)}$ over the
$k^{th}$ logical-qubit in quantum network under SCD.}\label{fig5}
\end{figure}

\begin{figure}[h]
\centering
\includegraphics[totalheight=2.0in]{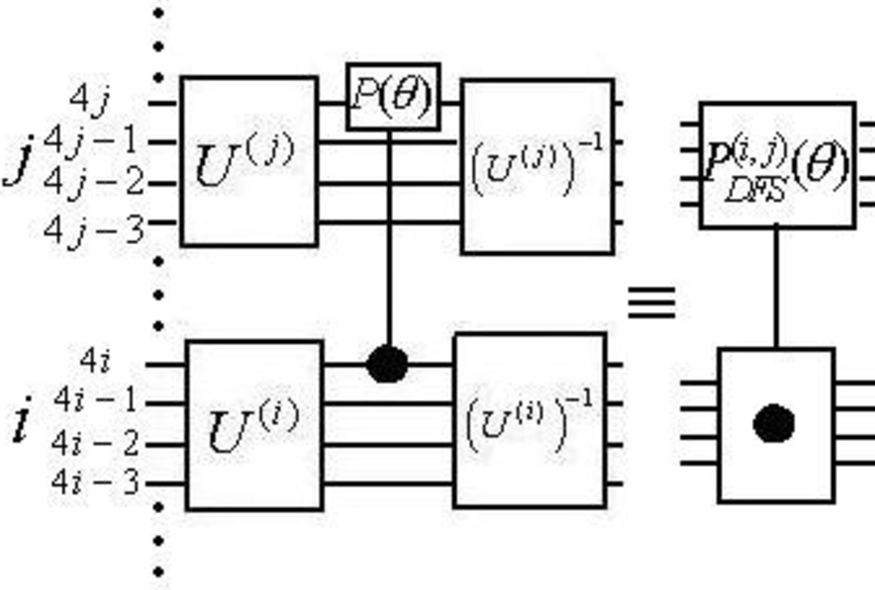}
\caption{Circuit for the controlled-phase-shift gate
$P_{DFS}^{(i,j)}(\theta)$ over the $i^{th}$ and $j^{th}$
logical-qubits in quantum network under SCD.} \label{fig6}
\end{figure}

With the help of the unitary transformation $U$, the fundamental
gates $H_{DFS}^{(k)}$ and $P_{DFS}^{(i,j)}(\theta)$  for
performing n-qubit QFT over the DFS of a 4n-qubit quantum network
under SCD are easy to be obtained(the corresponding circuits are
given in Fig.(\ref{fig5}) and Fig.(\ref{fig6}) respectively):

\noindent \begin{equation}\label{eq34}
H_{DFS}^{(k)}=U^{-1}H^{(4k)}U={U^{(k)}}^{-1}H^{(4k)}U^{(k)},
\end{equation}

\noindent\begin{eqnarray}\label{eq35}
P_{DFS}^{(i,j)}(\theta)&=&U^{-1}P^{(4i,4j)}(\theta)U \nonumber\\
&=&{U^{(i)}}^{-1}{U^{(j)}}^{-1}P^{(4i,4j)}(\theta)U^{(i)}U^{(j)},
\end{eqnarray}

\noindent where the gates $H_{DFS}^{(k)}$ and
$P_{DFS}^{(i,j)}(\theta)$ satisfy the requirements given section
{III}:

\noindent \begin{eqnarray}\label{eq36}
H_{DFS}^{(k)}|\tilde{0}\rangle_k&=&\frac{1}{\sqrt{2}}(|\tilde{0}\rangle_k+|\tilde{1}\rangle_k)\nonumber\\
H_{DFS}^{(k)}|\tilde{1}\rangle_k&=&\frac{1}{\sqrt{2}}(|\tilde{0}\rangle_k-|\tilde{1}\rangle_k)
\end{eqnarray}

\noindent and

\noindent \begin{eqnarray}\label{eq37}
&&P_{DFS}^{(i,j)}(\theta)|\tilde{s}_i\tilde{s}_j\rangle\nonumber\\
&=&\left\{\begin{array}{cc} |\tilde{s}_i\tilde{s}_j\rangle, & {\rm
if}|\tilde{s}_i\tilde{s}_j\rangle\in\{|\tilde{0}\tilde{0}\rangle,|\tilde{0}\tilde{1}\rangle,|\tilde{1}\tilde{0}\rangle\}\\
e^{i\theta}|\tilde{s}_i\tilde{s}_j\rangle, &{\rm if}\,\,
|\tilde{s}_i\tilde{s}_j\rangle=|\tilde{1}\tilde{1}\rangle
\end{array}\right.
.\end{eqnarray}

\noindent The circuit for performing QFT in the DFS of quantum
network is constructed by substituting the operators
$H_{DFS}^{(k)}$ and $P_{DFS}^{(i,j)}(\theta)$ into the gate
sequence in (\ref{eq13}).

\section{The efficiency and optimization}

The encoding efficiency of quantum algorithms over the DFS of an
$n$-qubit quantum network, say $\eta(n)$, is defined as the ratio
of the number of logical-qubits to that of physical-qubits. The
efficiency depends on the selection of DFS and the way of building
logical-qubits. From section {III}, it is obvious that the maximum
encoding efficiency is:

\noindent \begin{equation}\label{eq38}
\eta_{max}(n)=\frac{{\max}_J[{\log}_2(n_J)]}{n}.
\end{equation}

It has been derived in \cite{DFS3,DFS7} that the efficiency
$\eta_{max}(n)$ of the quantum network under collective
decoherence  approaches to 1 when $n\rightarrow\infty$. For the
circuits we designed for QFT over the quantum network under
collective decoherence, the encoding efficiency
$\eta(n)=\frac{1}{2}$ for WCD and $\eta(n)=\frac{1}{4}$ for SCD.
Therefore, it is possible to design a more efficient circuit for
realizing QFT over the DFS of some quantum network under
collective decoherence. However, our circuits are scalable for
they are relatively easy to be realized for large scale robust QFT
over quantum networks. Consequently, there is a trade-off between
the encoding efficiency and circuit complexity.

For example, if we want to implement $m$-qubit QFT in a DFS of
some quantum network under collective decoherence, then at least

\noindent \begin{equation}\label{eq39} r=\min
\{n|\max_J[\log_2(n_J)]\geq m\}
\end{equation}

\noindent physical-qubits are required. The corresponding circuit
for QFT over this $r$-qubit quantum network is the most efficient,
but it will become much more complicated in using more elementary
gates. The circuit design will be a formidable task.

\section{Conclusion}

In this paper, strategies for performing QFT in a quantum network
coupled with the environment are discussed. We propose a scheme
for noise-isolated QFT over the decoherence-free subspaces.
Following the scheme, circuits for implementing QFT are designed
in quantum network under collective decoherence. The efficiency of
these circuits and some possible improvements are discussed as
well.

In the future, a general designing methodology needs to be found
for more efficient QFT over arbitrary quantum network. Also, it is
worthwhile to reduce the number of elementary gates using in the
relevant quantum circuits. Moreover, it is interesting and useful
to extend the problem from the decoherence-free subspaces to
decoherence-free subsystems.

\begin{acknowledgments}
This work was supported by the NSFC-funded project No.60274025.
\end{acknowledgments}

\end{document}